%% file: artp174.tex
\input{psfig}

\documentstyle[prl,aps,twocolumn]{revtex}
\input{title2.tex}

\begin{document}

\title{\vspace*{-0.5cm}
\hspace*{\fill}{\normalsize LA-UR-98-2511} \\[1.5ex]
Predictions for $\mathbf{\sqrt{s}}$ = 200 A$\mathbf{\cdot}$GeV Au+Au Collisions
from Relativistic Hydrodynamics
}
\author{
B.R. Schlei${}^{1,2}$\thanks{E. Mail: schlei@LANL.gov}{\ } and
D. Strottman${}^2$\thanks{E. Mail: dds@LANL.gov}{\ }
\\[1.5ex]
{\it ${}^1$Physics Division, P-25, Los Alamos National Laboratory,
Los Alamos, NM 87545, USA\\
${}^2$Theoretical Division, DDT-DO, Los Alamos National Laboratory,
Los Alamos, NM 87545, USA}
}
\date{June 11, 1998}

\abstract{
The relativistic hydrodynamical model HYLANDER-C is used to give estimates for
single inclusive particle momentum spectra in $\sqrt{s}$ = 200
$GeV/nucleon$ Au+Au collisions that will be investigated experimentally in the
near future. The predictions are based on initial conditions that the initial
fireball has a longitudinal extension of 1.6 $fm$ and an initial energy density
of $30.8$ $GeV/fm^3$ as  obtained from a cascade model. For the collision
energy considered here, different stopping scenarios are explored for the first
time. Our calculations give particle yields of the order of 10,000 to 20,000
charged particles per event. }
\maketitle2
\narrowtext

Forthcoming experiments at the Relativistic Heavy-Ion Collider (RHIC) at
the Brookhaven National Laboratory (BNL) are designed to search for a new
state of nuclear matter, the quark-gluon plasma (QGP)
\cite{qm97}. The existence of a QGP should be reflected in the equation of
state (EOS) of nuclear matter, which is a relationship between extensive
variables, e.g.,  pressure, energy density, and baryon density.

Those theoretical models such as relativistic hydrodynamics which are
capable of describing relativistic heavy-ion collisions {\it and} which
also allow for an explicit use of an EOS have to be based on a
thermodynamical approach.  Several models  based on relativistic
hydrodynamics exist (for recent reviews {\it cf.}, e.g., Refs.
\cite{dan,trento97}); these models assume the existence of local
equilibrium. In general, the models solve the hydrodynamical equations --
which in the case of ideal relativistic fluids are given by the
relativistic Euler equations \cite{euler} -- by sophisticated numerical
techniques.

There are essentially only three numerical techniques that are employed in
the study of heavy-ion reactions \cite{dan}: particle methods, continuum
methods, and predictor-corrector methods. In our attempt to make
predictions for single inclusive particle momentum spectra of various
hadron species emerging from $\sqrt{s}$ = 200 $GeV/nucleon$ Au+Au
collisions, we shall use a predictor-corrector model, which has been
successful in providing a consistent and coherent description of the
single and double inclusive cross sections of mesons and baryons for
several types of fixed target heavy-ion collisions at CERN/SPS beam
energies near 200 A$\cdot$GeV: HYLANDER-C \cite{bernd15}, which is the
upgraded  successor of HYLANDER \cite{udo}.

Predictor-corrector type fluid models begin the calculations with the
matter already highly compressed and the numerical code then follows the
expansion of the fluid as governed by a particular EOS. At relativistic
energies, it is important to employ a covariant description of freeze-out,
which allows the calculation of the various particle spectra \cite{cooper}.
In earlier work, the initial state of the hot and dense zone of nuclear
matter -- the fireball -- was parametrized, and was adjusted until a match
with observables was established \cite{bernd_eos}. HYLANDER-C, e.g., can
use initial conditions ranging between the extremes defined by the Landau
\cite{landau} (i.e., complete stopping) and the Bjorken \cite{bjorken}
(i.e., minimum stopping through a scaling ansatz) initial conditions.

Conversely, one could also take information about the initial state of the
fireball from other sources, e.g., a cascade code, as input and then
allow the system to evolve according to an  assumed EOS, and finally
particle momentum spectra will be calculated. As there are no data yet
available for $\sqrt{s}$ = 200 $GeV/nucleon$ Au+Au collisions, we shall
discuss in this paper results for a relativistic heavy-ion
collision at these high energies using HYLANDER-C, and the sensitivity of
observables to different assumptions regarding different initial conditions.

For the following we have to specify an equation of state
and the initial conditions. In addition we choose for our freeze-out
condition a fixed  freeze-out energy density, $\epsilon_f$, and we assume
that the freeze-out  occurs for all particle species at the same fixed
value of $\epsilon_f$.  The calculations using HYLANDER and HYLANDER-C which
reproduced data of the CERN NA35, NA44 and NA49 collaborations
\cite{wenig} - \cite{NA49jones} were obtained while using an EOS with
a phase transition to a QGP at a critical  temperature $T_C=200\:MeV$ ({\it
cf.}  Refs. \cite{redlich,bernd8}, and Refs. therein). The EOS used in this
work does not depend on the baryon density (which is a valid assumption for
the collision energy regime considered here), and thus the  freeze-out
energy density translates into a fixed freeze-out temperature
$T_f$. The choice for the freeze-out temperature has been $T_f=139 MeV$ in
the former calculations; we shall use the same value here.  This is
consistent with our attempt to be conservative in determining possible
results for the Au+Au collisions by choosing exactly the same conditions
for the fireball expansion (through the same choice of  the EOS) and the
same conditions for the freeze-out (through the same choice  for
$\epsilon_f$ and $T_f$) that were used in the successful description of
the CERN/SPS data.

Because the appropriate initial conditions for the formation of the Au+Au
fireball are unknown, we shall adopt values for the hydrodynamical input
from K. Geiger's parton cascade calculations
\cite{geiger1,geiger2,geiger3}  (other cascade models \cite{wang1,wang2}
have been employed for predictions of the  heavy-ion collision discussed
here with results that are similar to K. Geiger's results). In Ref.
\cite{geiger2} it was reported that after an equilibration proper time,
$\tau_{eq}$ = 1.8 $fm/c$, the initial fireball originating from a
$\sqrt{s}$ = 200 $GeV/nucleon$ Au+Au collision was formed with an
initial longitudinal extension (of the Landau volume
\cite{jan}),  $\Delta$ = 1.6 $fm$, and an upper value for the initial energy
density,  $\epsilon_\Delta$ = 30.8 $GeV/fm^3$.

We use here for the initial distributions the initial condition scenario
which has been described in Refs. \cite{jan,bernd3}. Specifically, our
model uses the five initial parameters,
$K_L$, $\Delta$, $y_\Delta$, $y_m$ and $\sigma$, which are the relative
fraction of thermal energy in the central fireball, the longitudinal
extension of the fireball, the fluid's rapidity at the edge of the central
fireball, the fluid's rapidity at the maximum of the initial rapidity
distribution of the baryons, and the width of the initial baryonic rapidity
distribution, respectively. Furthermore, it is assumed that an initial
transverse fluid velocity component is absent, and the initial longitudinal
distributions (with respect to the  beam axis) for energy density,
$\epsilon$, and baryon density, $n_B$, are smeared out with a Woods-Saxon
parametrization in the transverse direction, $r_\perp$ ({\it cf.} Refs.
\cite{bernd3,sollfrank}).

In the following, we keep $\Delta$ and $K_L$ fixed to the values provided
by K. Geiger's results. We stress that an initial energy density,
$\epsilon_\Delta$ = 30.8 $GeV/fm^3$ in the Landau volume represents only
11.5\% of the total available energy.
The parameter $\sigma$ we keep fixed  also without loss of generality.
In particular, it is our intent to discuss six possible stopping scenarios
({\it cf.} Table I, scenarios I - VI). We can accomplish this by varying
the only two parameters left in our model for the initial conditions,
$y_\Delta$ and $y_m$. A larger value for the absolute value of  rapidity,
$y_\Delta$, at $z=\pm\Delta/2$ results in a larger initial rapidity field,
$y(z)$, which can be interpreted in having less stopping. Conversely,
initially concentrating as much baryonic matter as possible in the central
region of the initial fireball as energy conservation  permits us, i.e., by
minimizing $y_m$, results in having a much larger stopping compared to the
cases where the baryonic matter lies  further outside. For the following
discussion, we have chosen three  scenarios where as much baryonic matter as
possible is concentrated in the  central region of the initial fireball
(scenarios I, III,  V), and three  scenarios where the baryonic matter lies
further outside  (scenarios II, IV, VI).

In Fig. 1 we show the initial distributions of energy density,
$\epsilon$, baryon density, $n_B$, and the fluid rapidity, $y_F$,
normalized to their maximum values, $\epsilon^{max}$, $B^0_{max}$, and
$y_{cm}$, and plotted against the longitudinal coordinate $z$, for each of
the six initial conditions. The absolute values of the initial fluid
rapidity, $y_\Delta$, at $z=\pm\Delta/2$ decrease from the top to the bottom
of Fig. 1.  Therefore, we have a greater amount of stopping as one goes
from the top to the bottom of Fig. 1. Comparing the left column of Fig. 1
(scenarios I, III, V) with the right column (scenarios II, IV, VI), we
recognize that we have greater stopping when going from the right to the
left in the figure. Table I gives also the relative fractions,
$f_{n_B}^\Delta$, of baryons initially concentrated in the initial fireball
volume of length $\Delta$. Thus, scenario V gives the highest initial
stopping, and scenario II gives the smallest initial stopping.

Inspecting Fig. 1 further, we also see that the initial conditions I and
II are closer to a Bjorken-type initial condition, and scenarios V and VI
are closer to a Landau-type initial condition.

After fixing the various initial conditions, we allow for expansion and
cooling of the relativistic fluids by solving the relativistic Euler
equations. As mentioned above, freeze-out occurs for all particle species
at the same fixed value of $T_f=139\:MeV$. In Fig. 2 we display for each
scenario the isotherms of the expanding systems at the radial coordinate,
$r_\perp = 0$. In particular, the outer isotherms represent the  freeze-out
hypersurfaces at $r_\perp = 0$. The total lifetimes, $t_{max}$, of the
various systems range from 37.8 $fm/c$ to 49.7 $fm/c$, and the total
lifetimes of the QGP, $t_{QGP}$, range from 16.9 $fm/c$ to 23.7 $fm/c$. We
note that $t_{QGP}$ is not correlated with $t_{max}$ because of the complex
behaviour of the numerically treated multi-dimensional relativistic fluids.
Furthermore, looking at Fig. 2, we can see that the freeze-out
hypersurfaces are to a large extent hyperbola-like shaped, which we
especially would expect from a purely Bjorken type expansion scenario
\cite{bjorken}.

Following the formalism which is outlined in refs. \cite{jan,bernd3}, we
have calculated single inclusive particle momentum spectra of pions,
kaons, and protons for $\sqrt{s}$ = 200 $GeV/nucleon$  Au+Au collisions. We
have taken into account resonance decay  contributions up to the third
generation of particle production. Figs. 3 and 4 show our results for
rapidity and transverse momentum spectra, respectively, for the six initial
conditions. It is remarkable that a wide range of shapes is possible for
the rapidity spectra.  We observe an almost Gaussian distribution for
scenario V, a flat rapidity plateau for scenario III, and saddle-shaped
distributions for all the other four scenarios. In agreement with
expectation, the rapidity spectra are narrower and the transverse momentum
spectra show a higher mean transverse momentum,  $\langle k_\perp \rangle$,
(i.e., a smaller slope,) for initially higher stopping. Contrary to the
common belief \cite{bjorken}, Bjorken type expansion scenarios do not
necessarily lead to flat rapidity plateaus. This is because of finite size
effects of the fireballs.

In addition, we list in Table II the absolute particle yields for some
selected particle species. The numbers were obtained by integrating the
corresponding rapidity and transverse momentum spectra, respectively.
We stress that the particle yields in Table II are obtained while
assuming that $\epsilon_\Delta$ = 30.8 $GeV/fm^3$, i.e., $K_L$ = 11.5\%.
If one would choose pure Landau initial conditions (i.e., complete stopping)
one would obtain approximately 10 times that many particles.

To summarize, we have used the relativistic hydrodynamical model HYLANDER-C
to give estimates for single inclusive particle momentum spectra of various
hadrons for $\sqrt{s}$ = 200 $GeV/nucleon$ Au+Au collisions. Our
predictions are based on the assumptions that the initial fireball has a
longitudinal extension of 1.6 $fm$ and an initial energy density of $30.8$
$GeV/fm^3$, values consistent with results obtained using the cascade code
of Geiger. Within different stopping scenarios,  we have investigated the
freeze-out hypersurfaces and the corresponding final particle
distributions. Our calculations give particle yields of the order of 10,000
to 20,000 charged particles per event. We have also found that the shapes
of the rapidity spectra are sensitive to the assumed stopping power. Our
quantitative estimates are based on assumptions for the initial conditions
obtained from cascade models, the EOS, and the proper freeze-out conditions
for the heretofore experimentally unexplored regime of
$\sqrt{s}$ = 200 $GeV/nucleon$ collisions.  Information concerning each of
these must be gained with additional theoretical work as well as data when
they become available.

We are grateful for many instructive discussions with  Dr. Klaus
Kinder-Geiger. This work has been supported by the U.S. Department of
Energy.

\begin{table}
\vspace{-0.1cm}
\caption{Properties of the initial fireballs.}
\begin{center}
\begin{tabular}{l c c c c c c}
 & I & II & III & IV & V & VI \\
\hline
\multicolumn{7}{c}{Initial parameters}\\
$K_L$ & 0.115 & 0.115 & 0.115 & 0.115 & 0.115 & 0.115 \\
$\Delta$ $[fm]$ & 1.6 & 1.6 & 1.6 & 1.6 & 1.6 & 1.6 \\
$y_\Delta$ & 0.9 & 0.9 & 0.6 & 0.6 & 0.3 & 0.3 \\
$y_m$ & 1.52 & 2.00 & 1.24 & 1.75 & 0.88 & 1.50 \\
$\sigma$ & 0.4 & 0.4 & 0.4 & 0.4 & 0.4 & 0.4 \\
$T_f$ $[MeV]$ & 139 & 139 & 139 & 139 & 139 & 139 \\
\multicolumn{7}{c}{Output}\\
$y_{cm}$ & 5.36 & 5.36 & 5.36 & 5.36 & 5.36 & 5.36 \\
$\epsilon_\Delta$ $[GeV/fm^3]$ & 30.8 & 30.8 & 30.8 & 30.8 & 30.8 & 30.8 \\
$\epsilon^{max}$ $[GeV/fm^3]$ & 40.7 & 30.8 & 45.8 & 30.8 & 41.8 & 30.8 \\
$B^0_{max}$ $[fm^{-3}]$ & 2.23 & 2.09 & 1.52 & 1.44 & 0.78 & 0.74 \\
$f_{n_B}^\Delta$ & 0.062 & 0.003 & 0.055 & 0.002 & 0.072 & 0.001 \\
$t_{max}$ $[fm/c]$ & 49.7 & 51.7 & 45.6 & 49.0 & 37.8 & 43.5 \\
$t_{QGP}$ $[fm/c]$ & 23.7 & 16.9 & 23.5 & 20.3 & 21.2 & 20.6 \\
\end{tabular}
\end{center}
\end{table}

\begin{table}
\vspace{-0.7cm}
\caption{Particle yields of pions, kaons, and protons.}
\begin{center}
\begin{tabular}{l c c c c c c }
 & I & II & III & IV & V & VI \\
\hline
$N_{\pi^+}$ & 5320 & 4654 & 5802 & 5501 & 6816 & 5741 \\
$N_{\pi^-}$ & 5420 & 4748 & 5903 & 5609 & 6932 & 5840 \\
$N_{K^+}$ & 770 & 674 & 834 & 798 & 972 & 822 \\
$N_{K^-}$ & 709 & 616 & 776 & 735 & 915 & 766 \\
$N_p$ & 328 & 305 & 336 & 352 & 364 & 333 \\
\end{tabular}
\end{center}
\end{table}

\cleardoublepage

\vspace*{-2.0cm}
\begin{figure}
\begin{center}\mbox{ }
\[
\psfig{figure=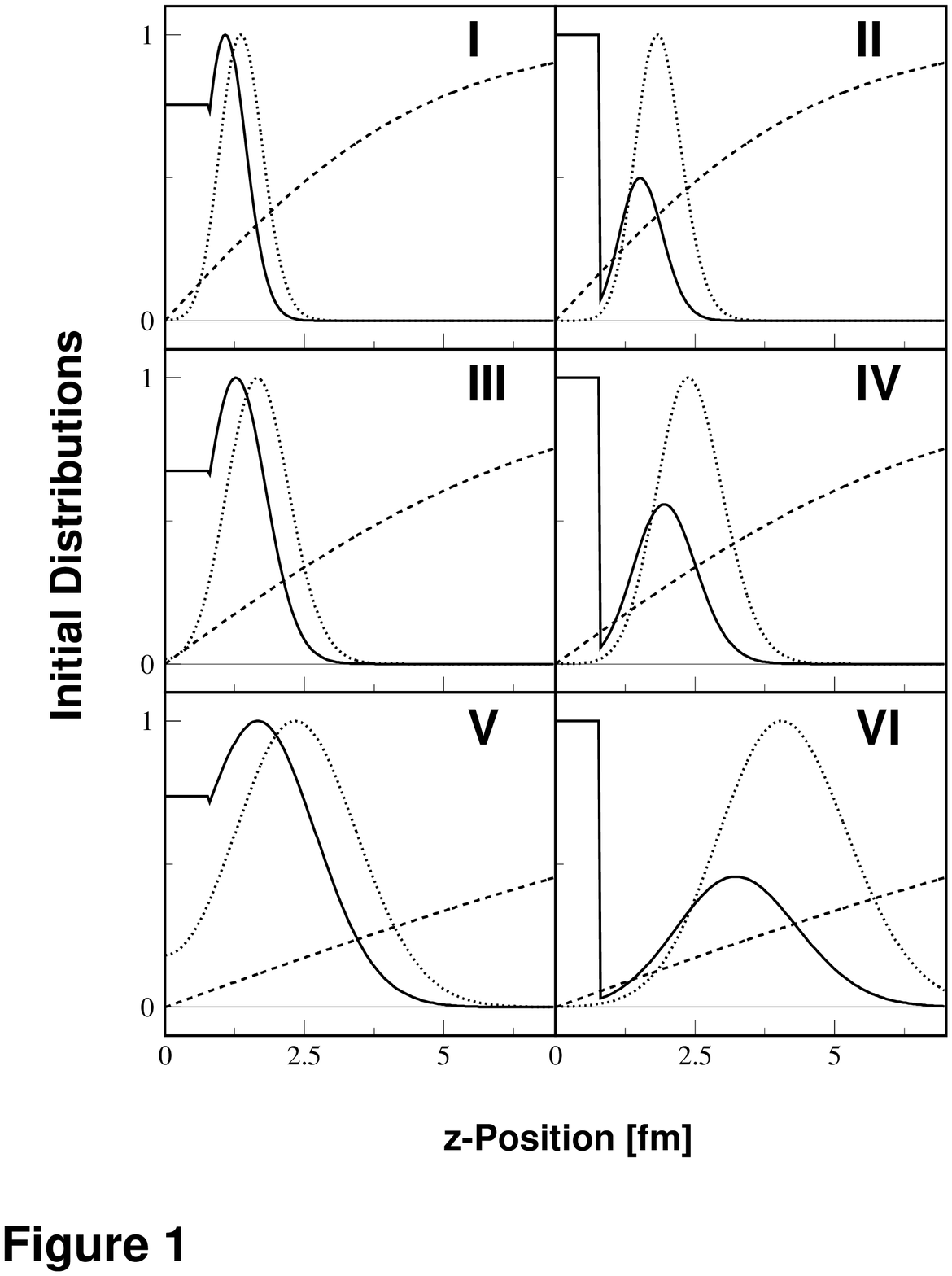,bbllx=0.0cm,bblly=3.0cm,%
bburx=21.0cm,bbury=29.7cm,width=8.5cm,clip=}
\]
\end{center}
\vspace*{-0.2cm}
\caption{Initial distributions of energy density, $\epsilon$
(solid lines), and baryon density, $n_B$ (dotted lines), as
well as the fluid rapidity, $y_F$ (dashed lines), normalized
to their maximum values (cf. Table I) and plotted against the
longitudinal coordinate $z$.}
\vspace*{-3.0cm}
\label{fg:fig1}
\end{figure}

\begin{figure}
\begin{center}\mbox{ }
\[
\psfig{figure=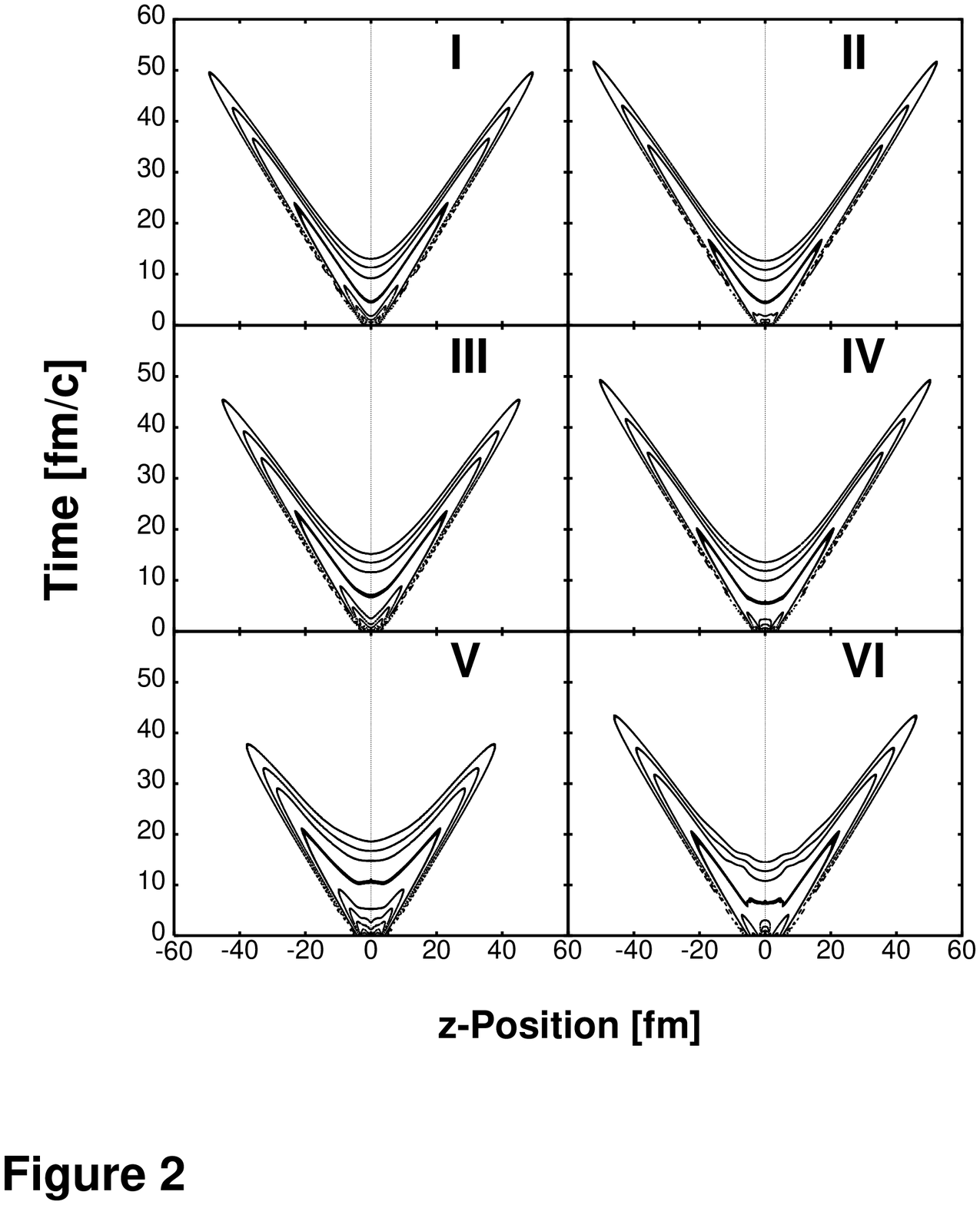,bbllx=0.0cm,bblly=4.0cm,%
bburx=21.0cm,bbury=29.7cm,width=8.5cm,clip=}
\]
\end{center}
\vspace*{-0.2cm}
\caption{Thermal evolution of the Au+Au fireballs.
In each case the outer contours are isotherms for
$T$ = 140 $MeV$, and each successively smaller
contour represents an increase of 20 $MeV$ in the
temperature.}
\label{fg:fig2}
\end{figure}

\newpage
\vspace*{-2.0cm}
\begin{figure}
\begin{center}\mbox{ }
\[
\psfig{figure=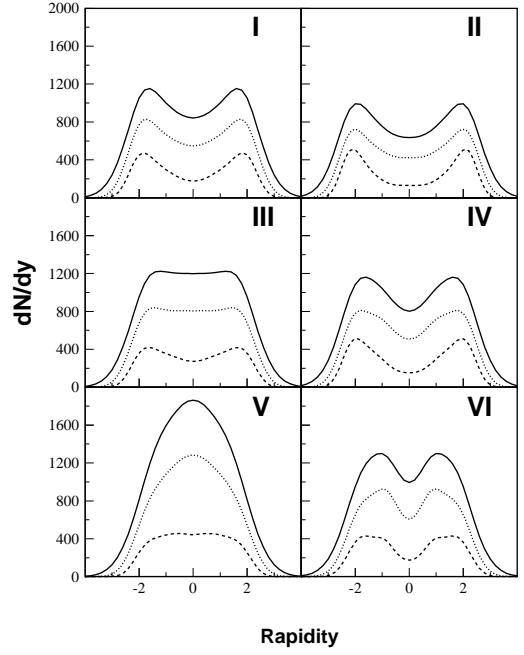,bbllx=0.0cm,bblly=3.0cm,%
bburx=21.0cm,bbury=29.7cm,width=8.5cm,clip=}
\]
\end{center}
\vspace*{-0.2cm}
\caption{Rapidity spectra of negative pions, $\pi^-$ (solid lines),
negative kaons, $K^-$ (dotted lines), and protons, $p$ (dashed
lines). In each plot the rapidity spectra of $K^-$ and $p$ are
enhanced by a factor of 5 for better visibility.}
\vspace*{-3.0cm}
\label{fg:fig3}
\end{figure}

\begin{figure}
\begin{center}\mbox{ }
\[
\psfig{figure=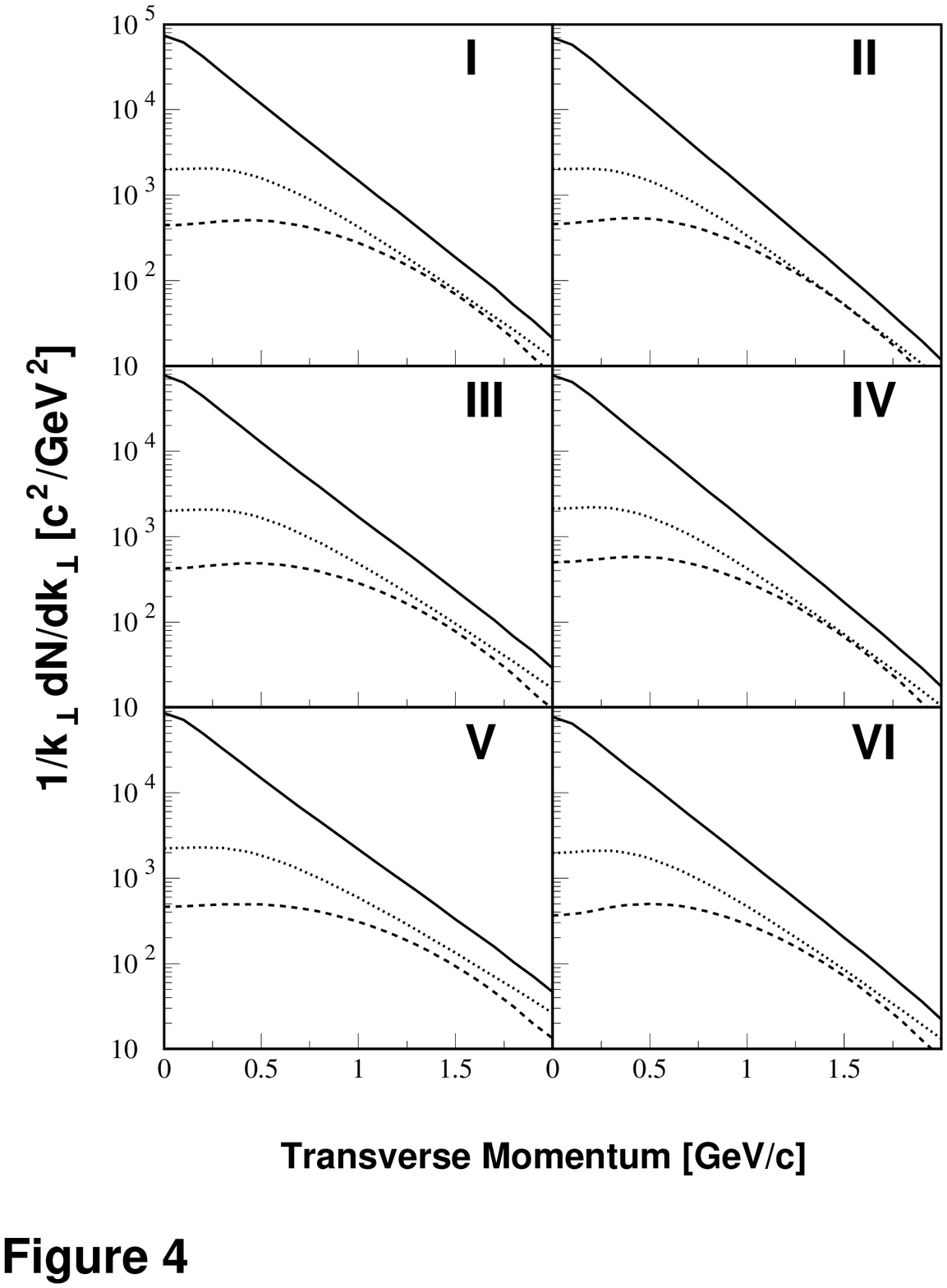,bbllx=0.0cm,bblly=3.0cm,%
bburx=21.0cm,bbury=29.7cm,width=8.5cm,clip=}
\]
\end{center}
\vspace*{-0.2cm}
\caption{Transverse momentum spectra of negative pions, $\pi^-$
(solid lines), negative kaons, $K^-$ (dotted lines), and protons,
$p$ (dashed lines).}
\label{fg:fig4}
\end{figure}


\end{document}

%% file: title2.tex
\catcode`\@=11

\def\maketitle2{\par 
\begingroup
\let\cite\@bylinecite
\def\thefootnote{\fnsymbol{footnote}}%
\twocolumn[\@maketitle2\vskip2pc]%
\thispagestyle{plain}\@thanks
\endgroup
\def\thefootnote{\arabic{footnote}}%
\setcounter{footnote}{0}%
\let\maketitle2\relax \let\@maketitle2\relax
\let\@thanks\relax \let\@authoraddress\relax \let\@title\relax
\let\@date\relax \let\thanks\relax \let\@abstract\relax 
\let\@pacs\relax}

\def\abstract#1{\gdef\@abstract{{\par 
\bgroup
\ifdim\prevdepth=-1000pt \prevdepth0pt\fi
\hsize\columnwidth
\dimen0=-\prevdepth \advance\dimen0 by17.5pt \nointerlineskip
\small\vrule width 0pt height\dimen0 \relax}{~~}#1\egroup}}

\def\pacs#1{\gdef\@pacs{{\par 
\bgroup
\hsize\columnwidth \parindent0pt
\ifdim\prevdepth=-1000pt \prevdepth0pt\fi
\dimen0=-\prevdepth \advance\dimen0 by20pt\nointerlineskip
\egroup} PACS numbers:~#1}}

\def\@maketitle2{
\@preprint
\@title
\ifdim\prevdepth=-1000pt \prevdepth0pt\fi
\@authoraddress
\@date
\begin{list}{}{\leftmargin=0.10753\textwidth \rightmargin=\leftmargin
\itemsep=1pc\partopsep=-1pc}
\item\@abstract
\item\@pacs
\end{list}
}

\catcode`\@=12